\documentclass[showpacs,amssymb,aps,twocolumn]{revtex4}
\usepackage{footmisc}
\usepackage{amsmath}
\usepackage{amstext}
\usepackage{amsopn}
\usepackage{amsfonts}
\usepackage{amssymb}
\usepackage{bbm}
\usepackage{accents}
\usepackage{empheq}
\usepackage{graphicx}
\usepackage{epsf}
\usepackage{graphics}
\usepackage[latin1]{inputenc}
\def\sc{\scriptscriptstyle}
\renewcommand{\thefootnote}{\fnsymbol{footnote}}
\long\def\symbolfootnote[#1]#2{\begingroup%
\def\thefootnote{\fnsymbol{footnote}}\footnote[#1]{#2}\endgroup}

\begin{document}
\title{An alternative construction of the positive inner product for pseudo-Hermitian Hamiltonians: 
Examples}

\author{Ashok Das$^{a,b}$, L. Greenwood$^{a}$\footnote{$\ $  e-mail: das@pas.rochester.edu,  
lgreenwo@pas.rochester.edu}}
\affiliation{$^a$ Department of Physics and Astronomy, University of Rochester, Rochester, NY 
14627-0171, USA}
\affiliation{$^b$ Saha Institute of Nuclear Physics, 1/AF Bidhannagar, Calcutta 700064, India}

\begin{abstract}
This paper builds on our earlier proposal for construction of a positive inner product for 
pseudo-Hermitian Hamiltonians and we give several examples to clarify our method. We show 
through the example of the harmonic oscillator how our construction applies equally well to 
Hermitian Hamiltonians which form a subset of pseudo-Hermitian systems. For  finite dimensional 
pseudo-Hermitian matrix Hamiltonians we construct the positive inner product (in the case of 
$2\times 2$ matrices for both real as well as complex eigenvalues). When the quantum mechanical 
system cannot be diagonalized exactly, our construction can be carried out perturbatively 
and we develop the general formalism for such a perturbative calculation systematically (for real 
eigenvalues). We illustrate how this general formalism works out in practice by calculating the inner 
product for a couple of ${\cal PT}$ symmetric quantum mechanical theories.  
\end{abstract}


\maketitle

\bigskip

\section{Introduction}

In a recent paper \cite{levi} we described a systematic procedure for constructing the positive inner 
product for a quantum mechanical systems described by a pseudo-Hermitian Hamiltonian which 
satisfies \cite{mo}
\begin{equation}
H = S^{-1} H^{\dagger} S.\label{pseudohermitian}
\end{equation}
Here $S$ is a bounded operator which can be chosen to be Hermitian (in the sense of Dirac). The 
crucial concept in our construction is the knowledge of the generators of energy eigenstates which 
acting on a given reference state generate all the eigenstates of the Hamiltonian. The method 
works equally well for systems with real or complex energy eigenvalues. We note that a 
pseudo-Hermitian Hamiltonian reduces to a Hermitian Hamiltonian when $S=\mathbbm{1}$ and, 
therefore, our construction of the positive inner product also reduces to the standard Dirac inner 
product in this case (which we discuss in section {\bf III}). In \cite{levi} we illustrated the method 
through the example of the Lee model \cite{tdlee,kallen} with an imaginary coupling constant 
\cite{Bender}. In this paper we give additional examples within the context of finite dimensional 
matrix Hamiltonians (which are pseudo-Hermitian) to clarify our method.

However, most quantum mechanical systems cannot be solved exactly. As a result, it is not possible 
to determine the energy eigenstates of the system and, therefore, their generators in a closed form. 
As we observed in \cite{levi}, in such a case, the generator as well as the inner product can be 
determined only perturbatively. (The pseudo-Hermitian systems differ from the Hermitian systems in 
this respect, namely, the inner product of the system depends on the dynamics of the system in a 
nontrivial manner \cite{mo,bender,mostafazadeh,bender-brody}.) In this paper, we would like to 
develop the idea of a perturbative determination of the inner product in detail for real energy 
eigenvalues (the discussion for complex energy values is straightforward) and work out various $
{\cal PT}$ symmetric \cite{bender,mostafazadeh} examples to illustrate the method.

The paper is organized as follows. In section {\bf II}, we briefly recapitulate our proposal for 
constructing the positive inner product in a pseudo-Hermitian quantum mechanical system, 
clarifying some of the details not fully explained in \cite{levi}. In section {\bf III}, we apply our method 
to the case of the harmonic oscillator which can be taken as the unperturbed Hamiltonian in some of 
the examples we discuss. In section {\bf IV}, we work out in detail a 
pseudo-Hermitian $2\times 2$ matrix model with real and complex eigenvalues in an analogy with 
\cite{bender-berry}. We also discuss the method in the context of $n \times n$ matrix models. In 
section {\bf V} we 
describe the general formalism for the perturbative determination of the generator of energy 
eigenstates and, therefore, the perturbative construction of the positive inner product (for real 
energy values). In section {\bf VI}, we apply the general method to various ${\cal PT}$ symmetric 
quantum mechanical models with real energy eigenvalues.  We conclude with a brief summary in 
section {\bf VII}.

\section{Recapitulation of the construction of the inner product}

In this section, we briefly recapitulate the essential points discussed in \cite{levi} in the construction 
of the 
inner product in a pseudo-Hermitian quantum mechanical system described by 
\eqref{pseudohermitian}. We showed that if we define an operator
\begin{equation}
q = S A,\label{q}
\end{equation}
where $[A, H] = 0$, and choose $A$ properly, then the quadratic form  
\begin{equation}
\langle \phi|\psi\rangle_{q} = \langle \phi|q|\psi\rangle,\label{qinnerproduct}
\end{equation}
defines an inner product (on a suitably defined Hilbert space, see \cite{levi} for details) satisfying 
\begin{equation}
\langle \psi_{\sc E'}|\psi_{\sc E}\rangle_{q} = \delta_{{\sc E}\bar{\sc E}'},\label{normalizedproduct}
\end{equation} 
with a unitary time evolution. Here and in what follows a bar denotes complex conjugation. The 
operator $q$ can be systematically constructed from a knowledge of the generators of the energy 
eigenstates of the theory in the following way.

An operator $\sigma_{\sc E}$ satisfying the relation 
\begin{eqnarray}
H\sigma_{\sc E}=E\sigma_{\sc E}+ \sigma_{\sc E} k_{\sc E},\label{generator}
\end{eqnarray}
is defined to be a generator of the eigenstates of $H$ with eigenvalue $E$ provided 

\begin{description}
\item[$(i)$] there exists at least one vector $|\psi\rangle$, solving
\begin{equation}
k_{\sc E}|\psi\rangle=0,\quad {\rm for\ all\ }E\in{\rm spect}(H),\label{k}
\end{equation}
with $\sigma_{\sc E}|\psi\rangle\neq0$.
\item[$(ii)$] there exists at least one vector $|\phi\rangle$ solving
\begin{equation}
k_{\sc E}^\dagger|\phi\rangle=0,\quad  {\rm for\ all\ }E\in{\rm spect}(H),\label{kdagger}
\end{equation}
with $\langle  \psi|\phi\rangle\neq0$.  
\item[$(iii)$] $\sigma_{\sc E}$ has an inverse $\sigma_{\sc E}^{-1}$, at least acting on the reference  state $|\psi\rangle$, such that 
\begin{equation}
\sigma_{\sc E}^{-1}\sigma_{\sc E}|\psi\rangle=|\psi\rangle,\label{partialinverse}
\end{equation}
and an adjoint  inverse, $(\sigma_{\sc E}^{\dagger})^{-1} $, well-defined 
when acting on $|\phi\rangle$, with $(\sigma_{\sc E}^{\dagger})^{-1} \;|\phi\rangle\neq0$.
\end{description}

From \eqref{generator} and \eqref{k}, it follows that 
\begin{equation}
|\psi_{\sc E}\rangle=\sigma_{\sc E}|\psi\rangle,\label{psiE}
\end{equation}
is an eigenvector of $H$ with eigenvalue $E$. Similarly, from the (Dirac) adjoint of \eqref{generator} 
as well as using \eqref{kdagger}, we can show that   
\begin{equation}
|\phi_{\sc E}\rangle=(\sigma_{\sc E}^{\dagger})^{-1} \;|\phi\rangle,\label{phiEbar}
\end{equation} 
is an eigenvector of $H^\dagger$ with eigenvalue $\bar{E}$. Furthermore, if $P_{\sc E}$ denotes 
projection operator on to the energy eigenstate $|\psi_{\sc E}\rangle$ with energy $E$
\begin{equation}
P_{\sc E} |\psi_{\sc E'}\rangle = \delta_{\sc EE'}\, |\psi_{\sc E}\rangle,\label{projection}
\end{equation}
then we showed in \cite{levi} that the action of $q$  can be expressed as
\begin{equation}
q (\sigma_{\sc E} |\psi\rangle) = (\sigma_{\bar{\sc E}}^{\dagger})^{-1} |\phi\rangle.
\end{equation}
As a result $q$ has the operator form
\begin{equation}
q  = \sum_{E} (\sigma_{\bar{\sc E}}^{\dagger})^{-1}  q_{0} \sigma_{\sc E}^{-1} P_{\sc E},
\label{qformula}
\end{equation}
where we have identified
\begin{equation}
|\phi\rangle = q_{0} |\psi\rangle,\quad {\rm such\ that}\quad \langle\psi|\phi\rangle = \langle\psi|q_{0}|
\psi\rangle = 1.\label{qzero}
\end{equation}
We note that in practical calculations, it is sufficient to have a partial inverse $^{\protect{\footnotemark[1]}}$\footnotetext[1]{A valid partial inverse can be found by noting that the generators are not unique.  In fact, if $\sigma_{\sc E}$ is a generator for $H$ then $\tilde{\sigma}_{\sc E}=f(H)\sigma_{\sc E}$ is also a generator (with the same reference state $|\psi\rangle$).  For the harmonic oscillator we get that $(\sigma^{-1})^{\dagger}=1/a$ as the generator for $H^{\dagger}$, but we prefer to use instead $H^\dagger (\sigma^{-1})^{\dagger}=a^\dagger$.} $\sigma_{\sc E}^{-1}$ as we will demonstrate in the next section.

The construction \eqref{generator}-\eqref{qzero} is quite general. It holds for pseudo-Hermitian 
Hamiltonians with real or complex eigenvalues. It also works for Hermitian Hamiltonians (which is a 
subset of pseudo-Hermitian Hamiltonians) for which $S=\mathbbm{1}$ and in this case the operator 
$q$ can be chosen to be $q = \mathbbm{1}$ leading us back to the Dirac inner product. In \cite{levi}  
we applied this construction to calculate $q$ for the Lee model \cite{tdlee,kallen} with an imaginary 
coupling \cite{Bender}. The purpose of the present paper is to work out more examples of diverse 
nature in order to shed more light on this method. It was pointed out in \cite{levi} that when the exact 
eigenstates of the Hamiltonian are not known, the generators as well as $q$ can be constructed 
perturbatively and one of the goals of the present work is to describe systematically how such a 
perturbative calculation works. In the next section we apply our method to the case of the harmonic 
oscillator which can be taken as the zeroth order Hamiltonian in some of the subsequent examples 
where we determine $q$ perturbatively. The Hamiltonian for the harmonic oscillator is, of course, 
Hermitian and this would also show how our method leads back to the standard Dirac inner product 
in cases where the Hamiltonian is Hermitian.

\section{Harmonic oscillator}

Before we determine $q$ for finite dimensional matrix models and discuss how $q$ can be 
determined perturbatively in systems where the exact energy eigenstates are not known, let us 
describe how the construction of the last section works in the simple example of the harmonic 
oscillator. In this case, we expect the inner product to coincide with the standard Dirac inner product 
and, therefore, we expect to be able to show that we can choose $q=\mathbbm{1}$. 

The Hamiltonian for the harmonic oscillator can be written as
\begin{equation}
H_{\rm h.o.} = \frac{1}{2} (p^{2} + x^{2} -1) = a^{\dagger}a = H^{\dagger}_{\rm h.o.},
\label{harmonicoscillator}
\end{equation}
where $a, a^{\dagger}$ denote respectively the annihilation and the creation operators defined by
\begin{equation}
a = \frac{1}{\sqrt{2}} (x + i p),\quad a^{\dagger} = \frac{1}{\sqrt{2}} (x-ip),\label{aadagger}
\end{equation}
and we have subtracted out the zero point energy in \eqref{harmonicoscillator} for simplicity. Here 
we have also set $m=\omega=1=\hbar$ for simplicity. The energy eigenstates of the oscillator 
are well known and satisfy
\begin{equation}
H_{\rm h.o.} |\psi_{\sc E_{n}}\rangle = E_{n} |\psi_{\sc E_{n}}\rangle = n |\psi_{n}\rangle,\quad  n = 
0,1,2,\cdots.\label{h.o.energy}
\end{equation}
Therefore, we can write the normalized eigenstates as 
\begin{equation}
|\psi_{\sc E_{n}}\rangle = |\psi_{n}\rangle = \frac{(a^{\dagger})^{n}}{\sqrt{n!}}\, |\psi_{0}\rangle,\quad 
a|\psi_{0}\rangle = 0,\label{h.o.eigenstates}
\end{equation}
where $|\psi_{0}\rangle$ denotes the ground state of the system (commonly denoted by $|0\rangle
$) and the energy eigenstates satisfy the orthonormality relation (with respect to the Dirac inner 
product)
\begin{equation}
\langle \psi_{m}|\psi_{n}\rangle = \delta_{mn}.\label{h.o.orthonormality}
\end{equation} 

We see from \eqref{h.o.eigenstates} that we can identify the reference state with the ground state 
and the generator of the energy eigenstates as
\begin{equation}
|\psi\rangle = |\psi_{0}\rangle,\quad \sigma_{\sc E_{n}} = \sigma_{n} = \frac{(a^{\dagger})^{n}}
{\sqrt{n!}}.\label{h.o.generator}
\end{equation}
Furthermore, from the canonical commutation relation $[a,a^{\dagger}] = \mathbbm{1}$, it follows 
that
\begin{equation}
H_{\rm h.o.} \sigma_{n} = E_{n} \sigma_{n} + \sigma_{n} H_{\rm h.o.},
\end{equation}
where we have used $E_{n} = n$. Comparing this with \eqref{generator} we determine that in the 
present case
\begin{equation}
k_{\sc E_{n}} = k_{n} = H_{\rm h.o.},
\end{equation}
which indeed satisfies
\begin{equation}
k_{n} |\psi\rangle = H_{\rm h.o.} |\psi_{0}\rangle = 0.
\end{equation}
It also follows that for the present problem
\begin{equation}
|\phi\rangle = |\psi\rangle = |\psi_{0}\rangle,\quad q_{0} = \mathbbm{1}.\label{qzeroho}
\end{equation}

The partial inverse of the generator \eqref{partialinverse} can also be determined from the canonical 
commutation relation to be
\begin{equation}
\sigma_{n}^{-1} = \frac{a^{n}}{\sqrt{n!}},\quad \sigma_{n}^{-1} \sigma_{n} |\psi\rangle = |\psi\rangle.
\end{equation}
We emphasize here that only a partial inverse is necessary for our construction which allows us to 
avoid awkward terms of the form $(a^{\dagger})^{-n}$ for the inverse. We also note that the 
projection operators for the energy eigenstates, in this case, can be written in the simple form
\begin{equation}
P_{\sc E_{n}} = P_{n} = |\psi_{n}\rangle \langle \psi_{n}|,\quad \sum_{n} P_{n} = \mathbbm{1},
\end{equation}
which satisfies
\begin{equation}
\frac{(a^{\dagger})^{n}}{\sqrt{n!}}\, \frac{a^{n}}{\sqrt{n!}}\, P_{n} = P_{n}.
\end{equation}
With all these, we can now determine $q$ (see \eqref{qformula}) to be
\begin{eqnarray}
q & = & \sum_{\sc E} (\sigma_{\sc E_{n}}^{-1})^{\dagger}  q_{0} \sigma_{\sc E_{n}}^{-1} P_{\sc E_{n}} 
=  \sum_{n} \frac{(a^{\dagger})^{n}}{\sqrt{n!}}\,\frac{a^{n}}{\sqrt{n!}}\, P_{n}\nonumber\\
& = & \sum_{n} P_{n} = \mathbbm{1},\label{qho}
\end{eqnarray}
so that the inner product \eqref{qinnerproduct}, in this case, coincides with the standard Dirac inner 
product
\begin{equation}
\langle \phi|\psi\rangle_{q} = \langle \phi|q|\psi\rangle = \langle \phi|\psi\rangle.
\end{equation} 

It is worth emphasizing here that the relation \eqref{generator} defining the generator does not 
determine its  scale uniquely, namely, the generator is defined only up to a multiplicative factor. If we 
had defined the generators in \eqref{h.o.generator} as $\sigma_{n} = c_{n} (a^{\dagger})^{n}$ 
instead (giving $\sigma_{n}^{-1} = \frac{a^{n}}{c_{n} n!}$), \eqref{qformula} would lead to 
\begin{equation}
q = \sum_{n} \frac{1}{|c_{n}|^{2} n!}\,P_{n}.
\end{equation}
Thus, we see that we can choose any coefficient multiplying the generators and it will change $q$ 
by a constant factor at each $P_{n}$. However, any choice of $\sigma_{n}$ enforces $
\langle \psi_{n}|\psi_{n}\rangle_{q} = 1$. In particular,  this determines that $q = \mathbbm{1}$ for 
$c_{n} = \frac{1}{\sqrt{n!}}$ reducing the inner product to the Dirac product.

\section{Pseudo-Hermitian matrix Hamiltonians}

In this section we discuss in detail two examples where we describe how our method applies to 
finite dimensional matrix Hamiltonians which are in general pseudo-Hermitian. The finite 
dimensional matrix Hamiltonians can be solved exactly in principle.

\subsection{$2\times 2$ matrix}

As the first example,  we apply our method to a simple $2\times 2$ matrix Hamiltonian with real as 
well as complex eigenvalues. We recall that the ${\cal PT}$ symmetric $2\times 2$ matrix model 
described by the Hamiltonian (see \cite{bender-berry} for details)
\begin{equation}
H^{({\cal PT})} = \begin{pmatrix}
r e^{i\theta} & s\\
s & r e^{-i\theta}
\end{pmatrix},\label{PT2x2}
\end{equation}
has been studied extensively in the past.  Let us, therefore,  analyze the generalized $2\times 2$ 
matrix Hamiltonain ($r,s,t,\theta,\phi$ are real parameters)
\begin{equation}
H = \begin{pmatrix}
r\, e^{i\theta} & s\, e^{i\phi}\\
t\, e^{-i\phi} & r\, e^{-i\theta}
\end{pmatrix},\label{general2x2}
\end{equation}
which is not Hermitian unless $s=t, \theta =0$. For $s=t, \phi=0$ this model reduces to the ${\cal PT}
$ symmetric theory \eqref{PT2x2}  where the parity operation is identified with
\begin{equation}
{\cal P} = \begin{pmatrix} 0 & 1\\
1 & 0
\end{pmatrix} = {\cal P}^{\dagger},\quad {\cal P}^{2} = \mathbbm{1},\label{2x2parity}
\end{equation}
with ${\cal T}$ denoting complex conjugation. The Hamiltonian in \eqref{general2x2}, however, is 
not ${\cal PT}$ symmetric in this context. On the other hand, if we define a ``generalized parity" 
operation 
through the $2\times 2$ matrix
\begin{equation}
\widetilde {\cal P} = \begin{pmatrix}
0 & \sqrt{\frac{s}{t}}\\
\sqrt{\frac{t}{s}} & 0
\end{pmatrix},\label{generalizedP}
\end{equation}
and choose time reversal to correspond to complex conjugation, then it is easily verified that the 
general Hamiltonian \eqref{general2x2} is $\widetilde {\cal P}{\cal T}$ symmetric, namely,
\begin{equation}
\widetilde {\cal P}{\cal T} H \widetilde {\cal P} {\cal T} = H.
\end{equation} 

The ``generalized parity" operator in \eqref{generalizedP} reduces to \eqref{2x2parity} when $s=t$. 
In general, however, let us note that while this operator is idempotent, it is not Hermitian,
\begin{equation}
\widetilde {\cal P} ^{2} = \mathbbm{1},\quad \widetilde {\cal P} ^{\dagger} \neq \widetilde {\cal P} .
\end{equation}
Furthermore, unlike the case in ${\cal PT}$ symmetric theories \cite{bender,mostafazadeh} (where $
{\cal P} H {\cal P} = H^{\dagger}$ and, therefore ${\cal P}$ defines the matrix $S$ in 
\eqref{pseudohermitian}), here we have
\begin{equation}
\widetilde {\cal P}  H \widetilde {\cal P}  = H^{*} \neq H^{\dagger}.
\end{equation}
Therefore, $\widetilde{\cal P}$ cannot be identified with $S$ in \eqref{pseudohermitian} which is 
used in our construction. However, we can define an operator ($2\times 2$ matrix)
\begin{equation}
S = \begin{pmatrix}
0 & e^{i\phi}\\
e^{-i\phi} & 0
\end{pmatrix} = S^{\dagger} = S^{-1},\label{2x2S}
\end{equation}
which leads to
\begin{equation}
H = S^{-1} H^{\dagger} S.
\end{equation}
In other words, even though $\widetilde {P} $ does not take the Hamiltonian to its Hermitian 
conjugate, $S$ does and the general Hamiltonian $H$ in \eqref{general2x2} is pseudo-Hermitian 
and our method can be directly applied.

The energy eigenvalues of this system are given by
\begin{equation}
E_{\pm} = r\,\cos\theta \pm \sqrt{st - r^{2} \sin^{2}\theta},\label{2x2evalues}
\end{equation}
and they are real for $st > r^{2}\,\sin^{2}\theta$ while they are complex for $st < r^{2}\,\sin^{2}\theta$. 
(We do not consider the degenerate case $st = r^{2}\,\sin^{2}\theta$, for which the Hamiltonian 
cannot be diagonalized, simply because it would introduce pseudo-eigenvectors.) We would 
analyze the two cases separately in the following.

\subsubsection{\bf Real eigenvalues:}

In the case of real eigenvalues, let us define
\begin{equation}
Q = \sqrt{st - r^{2}\,\sin^{2}\theta} = {\rm real},
\end{equation}
so that the two real energy eigenvalues can be written as
\begin{equation}
E_{\pm} = r\,\cos\theta \pm Q = \bar{E}_{\pm}.
\end{equation}
The corresponding eigenvalues can be determined to have the forms
\begin{eqnarray}
|\psi_{\sc E_{+}}\rangle \!\!&=& \!\frac{1}{\sqrt{s+t}}\!\begin{pmatrix}
\left(\frac{s}{t}\right)^{1/4}\,\sqrt{Q+ir\,\sin\theta}\,e^{i\phi/2}\\
\left(\frac{t}{s}\right)^{1/4}\,\sqrt{Q-ir\,\sin\theta}\,e^{-i\phi/2}
\end{pmatrix},\nonumber\\
\noalign{\vskip 4pt}%
|\psi_{\sc E_{-}}\rangle \!\!&=& \!\frac{i}{\sqrt{s+t}}\!\begin{pmatrix}
\left(\frac{s}{t}\right)^{1/4}\,\sqrt{Q-ir\,\sin\theta}\,e^{i\phi/2}\\
-\left(\frac{t}{s}\right)^{1/4}\,\sqrt{Q+ir\,\sin\theta}\,e^{-i\phi/2}
\!\!\end{pmatrix}\!\!,\label{realestates}
\end{eqnarray}
which have been normalized in the conventional sense for simplicity (although it is not necessary). 
We note that the two energy eigenstates are also eigenstates of $\widetilde{\cal P} {\cal T}$ with 
eigenvalue $1$ (they correspond to singlet states under $\widetilde{\cal P}{\cal T}$), namely,
\begin{equation}
\widetilde{\cal P} {\cal T} |\psi_{\sc E_{+}}\rangle = |\psi_{\sc E_{+}}\rangle,\quad \widetilde{\cal P} 
{\cal T} |\psi_{\sc E_{-}}\rangle = |\psi_{\sc E_{-}}\rangle.\label{pteigenstates}
\end{equation}
The projection operators onto the two states are given by
\begin{eqnarray}
P_{\sc E_{+}} & = & \frac{1}{2Q} \begin{pmatrix}
Q + ir\,\sin\theta & s\,e^{i\phi}\\
t\,e^{-i\phi} & Q - ir\,\sin\theta
\end{pmatrix},\nonumber\\
P_{\sc E_{-}} & = & \frac{1}{2Q} \begin{pmatrix}
Q - ir\,\sin\theta & -s\,e^{i\phi}\\
-t\,e^{-i\phi} & Q + ir\,\sin\theta
\end{pmatrix}.\label{realprojections}
\end{eqnarray}

Let us choose the reference state (see \eqref{k} or \eqref{psiE})
\begin{equation}
|\psi\rangle = |\psi_{\sc E_{-}}\rangle.
\end{equation}
It follows now from \eqref{realestates} that the generators of the states are given by 
\begin{eqnarray}
\sigma_{\sc E_{+}} & = & \begin{pmatrix}
0 & i \sqrt{\frac{s}{t}}\,e^{i\phi}\\
-i \sqrt{\frac{t}{s}}\,e^{-i\phi} & 0
\end{pmatrix} = \sigma_{\sc E_{+}}^{-1},\nonumber\\
\sigma_{\sc E_{-}} & = & \mathbbm{1},\quad k_{\sc E_{-}} = - k_{\sc E_{+}} = 2 Q P_{\sc E_{+}}.
\label{realgenerators}
\end{eqnarray}
The reference state \eqref{kdagger} can also be determined to be
\begin{equation}
|\phi\rangle = - \left(\frac{s+t}{2Q}\right) S|\psi\rangle = q_{0} |\psi\rangle,\quad q_{0} = -\left(\frac{s+t}
{2Q}\right) S,\label{realqzero}
\end{equation}
where the reference state has been normalized according to \eqref{qzero}. It follows now from 
\eqref{qformula} (using \eqref{realgenerators} as well as \eqref{realqzero}) that
\begin{eqnarray}
q & = & \sigma_{\sc E_{+}}^{\dagger} q_{0} \sigma_{\sc E_{+}} P_{\sc E_{+}} + \sigma_{\sc E_{-}} 
q_{0} \sigma_{\sc E_{-}} P_{\sc E_{-}}\nonumber\\
& = & \left(\frac{s+t}{2Q}\right) S \left(P_{\sc E_{+}} - P_{\sc E_{-}}\right)\nonumber\\
& = & \frac{s+t}{2Q^{2}} \begin{pmatrix}
t & - ir\,\sin\theta\,e^{i\phi}\\
ir\,\sin\theta\,e^{-i\phi} & s
\end{pmatrix}.\label{realq}
\end{eqnarray}
It can be checked now that
\begin{equation}
\langle\psi_{\sc E_{i}}|q|\psi_{\sc E_{j}}\rangle = \langle\psi_{\sc E_{i}}|\psi_{\sc E_{j}}\rangle_{q} = 
\delta_{ij},\quad i,j=\pm.
\end{equation}

\subsubsection{\bf Complex eigenvalues:}

For the case $st < r^{2}\,\sin^{2}\theta$, as we have noted the eigenvalues are complex and let us 
define
\begin{equation}
\widetilde {Q}  = \sqrt{r^{2}\,\sin^{2}\theta - st} = {\rm real},\label{Rbar}
\end{equation}
so that the two energy eigenvalues \eqref{2x2evalues} which become complex conjugate pairs can 
be written as
\begin{equation}
E = r\,\cos\theta - i \widetilde {Q} ,\quad \bar{E} = r\,\cos\theta + i \widetilde {Q} .
\label{complexevalues}
\end{equation}

The two eigenstates can be determined to have the forms
\begin{eqnarray}
|\psi_{\sc E}\rangle & = & \frac{(-i)}{\sqrt{(s+t)r\,\sin\theta + (s-t)\widetilde {Q}}}\nonumber\\
& & \times\begin{pmatrix}
i\sqrt{s(r\,\sin\theta - \widetilde {Q})}\,e^{i\phi/2}\\
\sqrt{t(r\,\sin\theta + \widetilde {Q})}\,e^{-i\phi/2}
\end{pmatrix},\nonumber\\
|\psi_{\sc \bar{E} }\rangle & = & \frac{i}{\sqrt{(s+t)r\,\sin\theta + (s-t)\widetilde {Q}}}\nonumber\\
& &  \times \begin{pmatrix}
\sqrt{s(r\,\sin\theta + \widetilde {Q})}\,e^{i\phi/2}\\
-i\sqrt{t(r\,\sin\theta - \widetilde {Q})}\,e^{-i\phi/2}
\end{pmatrix},\label{complexestates}
\end{eqnarray}
It s clear that the eigenstates \eqref{complexestates} of the Hamiltonian for complex eigenvalues 
correspond to a doublet representation of $\widetilde {\cal P}{\cal T}$ (compare with 
\eqref{pteigenstates}), 
namely,
\begin{equation}
\widetilde {\cal P} {\cal T} |\psi_{\sc E}\rangle = |\psi_{\sc \bar{E} }\rangle,\quad \widetilde {\cal P}{\cal 
T} |\psi_{\sc 
\bar{E} }\rangle = |\psi_{\sc E}\rangle.
\end{equation}
The projection operators onto these two states take the forms
\begin{eqnarray}
P_{\sc E} & = & \frac{1}{2\widetilde {Q} }\begin{pmatrix}
 - (r\,\sin\theta - \widetilde {Q}) &  is\,e^{i\phi}\\
it\,e^{-i\phi} &  r\,\sin\theta + \widetilde {Q}
\end{pmatrix},\nonumber\\
P_{\sc \bar{E} } & = & \frac{1}{2\widetilde {Q} }\begin{pmatrix}
r\,\sin\theta + \widetilde {Q} & -is\,e^{i\phi}\\
-it\,e^{-i\phi} &  - (r\,\sin\theta - \widetilde {Q})
\end{pmatrix}.\label{complexprojections}
\end{eqnarray}

Let us next choose the reference state
\begin{equation}
|\psi\rangle = |\psi_{\sc E}\rangle.
\end{equation}
It follows now from \eqref{complexestates} that the generators of states are given by 
\begin{eqnarray}
\sigma_{\sc \bar{E} } & = & \begin{pmatrix}
0 & - \sqrt{\frac{s}{t}}\,e^{i\phi}\\
\sqrt{\frac{t}{s}}\,e^{-i\phi} & 0
\end{pmatrix} = - \sigma_{\sc \bar{E} }^{-1},\nonumber\\
\sigma_{\sc E} & = & \mathbbm{1},\quad k_{\sc E} = - k_{\sc \bar{E} } = 2i\widetilde {Q}  P_{\sc 
\bar{E} }.\label{complexgenerators}
\end{eqnarray}
The reference state \eqref{kdagger} can now be obtained to have the form
\begin{equation}
|\phi\rangle = q_{0}|\psi\rangle= - \frac{(s+t) r\,\sin\theta + (s-t)\widetilde {Q}}{2 \sqrt{st}\,\widetilde{Q}}
\,\sigma_{\sc \bar{E} }^{\dagger} S |\psi\rangle,
\end{equation}
where
\begin{equation}
q_{0} = - \frac{(s+t) r\,\sin\theta + (s-t)\widetilde {Q}}{2 \sqrt{st}\,\widetilde{Q}}\begin{pmatrix}
\sqrt{\frac{t}{s}} & 0\\
0 & - \sqrt{\frac{s}{t}}
\end{pmatrix}.\label{complexqzero}
\end{equation}
It can be checked that this state satisfies
\begin{equation}
k_{\sc E}^{\dagger} |\phi\rangle =  k_{\sc \bar{E} }^{\dagger} |\phi\rangle = 0,\quad 
\langle \psi|\phi\rangle =  \langle\psi|q_{0}|\psi\rangle =1.
\end{equation}

It follows now from \eqref{qformula} (using \eqref{complexgenerators} as well as 
\eqref{complexqzero}) that
\begin{eqnarray}
q & = & - \sigma_{\sc \bar{E}}^{\dagger} q_{0} P_{\sc E} - q_{0}\sigma_{\sc \bar{E}} P_{\sc \bar{E}}
\nonumber\\
& = & - \frac{(s+t) r\,\sin\theta + (s-t)\widetilde {Q}}{2 \sqrt{st}\,\widetilde{Q}}\,S (P_{\sc E} + P_{\sc 
\bar{E}})\nonumber\\
& = & - \frac{(s+t) r\,\sin\theta + (s-t)\widetilde {Q}}{2 \sqrt{st}\,\widetilde{Q}}\,S.
\end{eqnarray}
It is easy to check that
\begin{equation}
\langle \psi_{\sc E}|\psi_{\sc E}\rangle_{q} = 0 = \langle\psi_{\sc \bar{E}}|\psi_{\sc \bar{E}}\rangle_{q},
\quad \langle\psi_{\sc \bar{E}}|\psi_{\sc E}\rangle_{q} = 1.
\end{equation}
In this case, we note that the operator $q$ simply scales the operator $S$ in \eqref{2x2S} which 
leads to the pseudo-Hermitian nature of the Hamiltonian. 

\subsection{$\mathbf{n}\times \mathbf{n}$ matrix}

Here we will solve for the operator $q$ for an $n$-dimensional Hilbert space, so that $q$ will be an 
$n\times n$ matrix.  We will assume that the Hamiltonian $H$ has $n$ real, distinct eigenvalues (no 
degeneracies) so that $H$ has a diagonal Jordan decomposition. Therefore, we can find an 
invertible matrix $R$ such that 
\begin{equation}
R^{-1}HR=H_0,
\end{equation}
where $H_0=\;$diag$\{E_1, E_2, ..., E_n\}$.  Because $H_0$ is diagonal there exists a basis of 
vectors, $|v_i\rangle$ satisfying, 
\begin{equation}
H_0|v_i\rangle=E_i|v_i\rangle,\quad \langle v_j|v_i\rangle=\delta_{ji},\quad E_{i} = \bar{E}_{i}, 
\end{equation}
for $1\leq i,j\leq n$.
 
However, in the non-diagonal basis, the eigenstates for $H$ are given by $|\psi_i\rangle=R|v_i
\rangle$.  The Dirac-inner product of two such states is  
\begin{equation}
\langle\psi_i|\psi_j\rangle = \langle v_i|R^\dagger R|v_j\rangle,\label{one}
\end{equation}
and is generally not orthonormal with respect to the eigenstates $|\psi_i\rangle$.  We can modify the 
inner product by introducing an operator $q$ as discussed in the previous section,
\begin{equation}
\langle\psi_i|\psi_j\rangle_q = \langle\psi_i|q|\psi_j\rangle=\delta_{ij}.\label{two}
\end{equation}
In order to restore orthonormality it would appear from \eqref{one} and \eqref{two} that we should 
choose $q= (R^{-1})^{\dagger} 
R^{-1}$. On the other hand, we recall from \eqref{qformula} that 
\begin{equation}
q=\sum_{i=1}^n (\sigma_{E_i}^{-1})^{\dagger}  q_0 \sigma_{E_i}^{-1}P_i,
\end{equation}
where $\sigma_{E_i}|\psi\rangle=|\psi_i\rangle$ defines the generators of states, $P_j|\psi_i\rangle=
\delta_{ij}|\psi_i\rangle$ defines the projection operators, and $q_0$ is the matrix satisfying $q_0|
\psi\rangle=|\phi\rangle, \langle\psi|\phi\rangle =1,$ for the reference states $|\psi\rangle,\ |\phi
\rangle$.  

To show that the two definitions are in fact equivalent,  let us choose the reference state to be  $|\psi
\rangle=|\psi_n\rangle$. Since $|\psi_n
\rangle=R|v_n\rangle$, this implies that $|\phi\rangle=(R^{-1})^{\dagger}  |v_n\rangle$ (because we 
want $\langle\phi|\psi\rangle=1$ to hold).  Furthermore, since  $q_0|\psi_n\rangle = q_{0} |\psi
\rangle =|\phi\rangle$, then using $P_n=R|v_n\rangle\langle v_n|R^{-1}=|\psi_n\rangle \langle 
\psi_n| q$ we can write the identity 
\begin{equation}
q_0P_n=qP_n.
\end{equation}
(We note that although $q_0$ is generally different from $q$, it must act the same way on the 
reference state $|\psi_n\rangle$.)

Second, we observe that in the diagonal basis, $|v_i\rangle$, there exists an upper-triangular matrix 
$t$ satisfying $t|v_i\rangle=|v_{i+1}\rangle$, with $t|v_n\rangle=|v_1\rangle$, and $t^
\dagger=t^{-1}$ ($t$ is just the matrix with ones above the diagonal, and one in the lower left-hand 
corner).  This gives $\sigma_{E_i}=\sigma^i=Rt^{i}R^{-1}$ (here $t^{i}$ denotes the $i$ th power of 
$t$), where $\sigma=RtR^{-1}$ defines the 
generator of states.  The statement $\sigma^{-i}P_i=P_n\sigma^{-i}$ is evident.

Using our definitions above, we can calculate directly 
\begin{eqnarray}
q &=& \sum_{i=1}^n (\sigma^{-i})^{\dagger} q_0\sigma^{-i}P_i\nonumber\\
&=&\sum_{i=1}^n (\sigma^{-i})^{\dagger}  q_0P_n
\sigma^{-i}=\sum_{i=1}^n (\sigma^{-i})^{\dagger}  qP_n\sigma^{-i}\nonumber\\
&=&\sum_{i=1}^n ((R^{-1})^{\dagger} t^{i}R^{\dagger})((R^{-1})^{\dagger} R^{-1})P_n (Rt^{-i}R^{-1})
\nonumber\\
&=& (R^{-1})^{\dagger} \left(\sum_{i=1}^n|v_i\rangle\langle v_i|\right)\ R^{-1}\nonumber\\
&=& (R^{-1})^{\dagger}  R^{-1}.
\end{eqnarray}

In finite dimensional Hilbert spaces, such as this, it is probably easier to calculate 
$q= (R^{-1})^{\dagger} R^{-1}$ directly from the Jordan matrices $R$ (which consist of each 
eigenvector 
of $H$ in one of the columns).  However, in infinite dimensional Hilbert spaces there is no Jordan 
decomposition, but the operators $\sigma_E$ which generate the eigenvectors do exist.  Thus, our 
expression for $q$ is valid in both finite, and infinite dimensions, whether or not one can construct 
the Jordan matrix.  As we will see in the next two sections, there exist operators (also 
denoted by $R$) which will play a role very similar to the Jordan matrix given above.  These will turn 
out to be the perturbation operators.

\section{Perturbative determination of $q$}

Let us next consider a pseudo-Hermitian system with real energy eigenvalues (as in ${\cal PT}$ 
symmetric theories \cite{bender,mostafazadeh}) where the energy eigenstates may be difficult to 
determine exactly. In this case, let us assume that we can write the pseudo-Hermitian Hamiltonian  
$H$ as 
\begin{eqnarray}
H = H_0+\epsilon V(x)\label{perturbation},
\end{eqnarray}
where $H_{0}$ is the part of the Hamiltonian $H$ which we can diagonalize and we treat $\epsilon 
V (x)$ as a perturbation. Of course, $H_{0}$ can be Hermitian or non-Hermitian. However, in most 
practical examples, it can be chosen to be Hermitian and this is the case we will discuss here. Let 
us denote the generators for the eigenstates of $H_0$ by $\sigma_{\sc E}^{(0)}$ so that we can 
express the eigenstates of $H_{0}$ as
\begin{eqnarray}
|\psi_{\sc E}^{(0)}\rangle=\sigma_{\sc E}^{(0)}|\psi\rangle,
\end{eqnarray}
where $|\psi\rangle$ denotes the reference state for the system (see \eqref{k} and \eqref{psiE}). 
Since $H_{0}$ is chosen to be Hermitian, as in the case of the harmonic oscillator, it follows that 
(see \eqref{qzeroho})
\begin{equation}
|\psi\rangle = |\phi\rangle,\quad q_{0} = \mathbbm{1}.\label{psiphi}
\end{equation}
 Thus, for the (diagonalizable) unperturbed part of the Hamiltonian we can determine (see 
\eqref{qho}) 
\begin{eqnarray}
q^{(0)} & = & \sum_{\sc E} (\sigma_{\sc E}^{(0)\; -1})^{\dagger}\; q_{0} (\sigma_{\sc 
E}^{(0)})^{-1}\;P_{\sc 
E}^{(0)}\nonumber\\
& = & \sum_{\sc E} (\sigma_{\sc E}^{(0)\; -1})^{\dagger}\, (\sigma_{\sc E}^{(0)})^{-1}\;P_{\sc 
E}^{(0)} = \mathbbm{1},\label{qsuperzero}
\end{eqnarray}
as we have seen in \eqref{qho}. Here $P^{(0)}_{\sc E}=|\psi_{\sc E}^{(0)}\rangle\langle  \psi_{\sc 
E}^{(0)}|$ is the projection onto an eigenstate of $H_0$ with eigenvalue $E^{(0)}$.    

The eigenstates of the total Hamiltonian $H$ in \eqref{perturbation} can be written as a series in 
powers of the perturbation parameter $\epsilon$, namely, ($\epsilon^{i}$ denotes the $i$th power 
of $\epsilon$)
\begin{eqnarray}
|\psi_{\sc E}\rangle = \sum_{i} \epsilon^{i} |\psi_{\sc E}^{(i)}\rangle,\label{perturbationseries}
\end{eqnarray} 
where $|\psi_{\sc E}^{(i)}\rangle$ denotes the $i$th order correction to the state $|\psi_{E}^{(0)}
\rangle$. We can always choose the corrections to a given state to be orthogonal to the state, 
namely,
\begin{equation}
\langle \psi_{\sc E}^{(0)}|\psi_{\sc E}^{(i)}\rangle = 0,\quad i>0.
\end{equation}
In this case, the perturbative corrections to the state $|\psi_{\sc E}^{(0)}\rangle$ can be obtained to 
have the compact recursive form ($i>0$ and we are assuming the energy eigenvalues to be 
discrete)
\begin{eqnarray}
|\psi_{\sc E}^{(i)}\rangle & = & \sum_{\sc E'\neq E} \frac{1}{E^{(0)} - E'^{(0)}}\times \nonumber\\
& &\!\!\!\left(\!\langle \psi_{\sc E'}^{(0)}|V|\psi_{\sc E}^{(i-1)}\rangle - \sum_{j=1}^{i-1} E^{(j)} \langle 
\psi_{\sc E'}^{(0)}|\psi_{\sc E}^{(i-j)}\rangle\!\right) \!|\psi_{\sc E'}^{(0)}\rangle\nonumber\\
& = & R^{(i)} |\psi_{\sc E}^{(0)}\rangle = R^{(i)} \sigma_{\sc E}^{(0)} |\psi\rangle = \sigma_{\sc E}^{(i)} |
\psi\rangle,\label{wfncorrection}
\end{eqnarray}
where $R^{(i)}$ denotes the operator that transforms the unperturbed state $|\psi_{\sc E}^{(0)}
\rangle
$ to $|\psi_{\sc E}^{(i)}\rangle$ and
\begin{equation}
E^{(i)} = \langle \psi_{\sc E}^{(0)}|V|\psi_{\sc E}^{(i-1)}\rangle,\quad i>0.
\end{equation}

From \eqref{wfncorrection} we note that we can define the correction to the generator of the state at 
the $i$th order as 
\begin{eqnarray}
\sigma_{\sc E}^{(i)} = R^{(i)} \sigma_{\sc E}^{(0)},\label{Ri}
\end{eqnarray}
with the same (zeroth order) reference state $|\psi\rangle$.  Therefore, to any order in perturbation 
we can write
\begin{eqnarray}
R & = & \sum_{i=0} \epsilon^{i} R^{(i)},\nonumber\\
\sigma_{\sc E} & = & \sum_{i=0} \epsilon^{i} \sigma_{\sc E}^{(i)} = \sum_{i=0}\epsilon^{i} R^{(i)} 
\sigma_{\sc E}^{(0)} = R \sigma_{\sc E}^{(0)},\label{R}
\end{eqnarray}
where $R^{(0)} = \mathbbm{1}$. By taking the adjoint of \eqref{perturbation}, and using $|\phi
\rangle=|\psi\rangle$ (see \eqref{psiphi}), we likewise determine the corrections to the eigenstates of 
$H^\dagger$ to 
any order to be (see \eqref{phiEbar})
\begin{eqnarray}
|\phi_{\sc E}\rangle=(R^{-1})^{\dagger} |\phi_{\sc E}^{(0)}\rangle=(R^{-1})^{\dagger} (\sigma_{\sc 
E}^{(0)\; -1})^{\dagger}\;|\psi\rangle,
\end{eqnarray}
with the same reference state $|\psi\rangle$.  

Puting this into the equation for $q$ (see \eqref{qformula}) we find that for real $E$ (recall that 
we are considering a pseudo-Hermitian system with real energy eigenvalues as in ${\cal PT}$ 
symmetric theories)
\begin{eqnarray}
q &=&\sum_{\sc E} (\sigma_{\sc E}^{-1})^{\dagger} \;\sigma_{\sc E}^{-1}\;P_{\sc E}\nonumber\\
&=&(R^{-1})^{\dagger} \left(\sum_{\sc E} (\sigma_{\sc E}^{(0)\; -1})^{\dagger}\;\sigma_{\sc 
E}^{(0)\;-1}\;P_{\sc E}^{(0)}\right)R^{-1}\nonumber\\
&=&(R^{-1})^{\dagger} \;q^{(0)}\;R^{-1},\label{nonhermitian}
\end{eqnarray}
where $q^{(0)}$ is the $q$ operator for the unperturbed Hamiltonian which, for a Hermitian $H_{0}$, 
is given by $q^{(0)}= \mathbbm{1}$ (see \eqref{qsuperzero}). In this case, therefore, 
\eqref{nonhermitian} reduces to
\begin{eqnarray}
q=(R^{-1})^{\dagger}  R^{-1}.\label{qR}
\end{eqnarray}
For real eigenvalues, we see that  $q$ only depends on the perturbation operator $R$, and that the 
problem is solved 
once we determine $R$ from  perturbation theory.  If we had not chosen $H_0$ to be Hermitian, 
then $q^{(0)}$ in   \eqref{nonhermitian} would be nontrivial which would need to be 
determined as well.

\section{Examples}

In this section we would work out two examples of ${\cal PT}$ symmetric theories (with real energy 
eigenvalues) to illustrate how the perturbative calculation is carried out in practice. We recall from 
\eqref{qR} that when the unperturbed Hamiltonian $H_{0}$ is Hermitian, then to any order in the 
perturbing parameter, the operator $q$ is determined from the operator $R$ defined in \eqref{Ri}-
\eqref{R} to that order.

\subsection{Example: $H = \frac{1}{2} (p^{2} + x^{2} -1) + i\epsilon x^{3}$}

Let us consider the Hamiltonian given by \cite{bender-meisinger},
\begin{eqnarray}
H = \frac{1}{2} (p^{2} + x^{2} -1) + i\epsilon x^{3} = H_{0} + \epsilon V (x),\label{cubic}
\end{eqnarray}
where $\epsilon$ is a real constant parameter and we recognize that $H_0 = H_{\rm h.o.}$ 
describes the harmonic oscillator which we have studied in section 
{\bf III} (we continue to identify $m=\omega = 1=\hbar$).  This model is known to be ${\cal PT}$ 
symmetric with real energy eigenvalues and has been studied extensively in the past. Here we 
would calculate the operator $q$ (which defines the inner product) associated with this system 
perturbatively using our method. 

The Hamiltonian for the harmonic oscillator is Hermitian and the system can be solved exactly. As a 
result, we can choose $H_{0}=H_{\rm h.o.}$ to be the unperturbed Hamiltonian in which case we 
can take over 
the analysis of section {\bf V}. We have already seen in \eqref{h.o.eigenstates} and 
\eqref{h.o.generator} that the generator for the harmonic oscillator has a very simple form. 
Therefore, for the unperturbed system we can identify
\begin{equation}
|\psi^{(0)}_n\rangle = \sigma_n^{(0)}|\psi_{0}\rangle = \sigma_{n}^{(0)}|\psi\rangle,\quad E_{n}^{(0)} 
= n.
\end{equation} 
Furthermore, recalling that $x=\frac{1}{\sqrt{2}}(a+a^\dagger)$ (see, \eqref{aadagger}), the first order 
correction to the unperturbed eigenstate can be determined from \eqref{wfncorrection} as 
\begin{eqnarray}
|\psi^{(1)}_n\rangle&=&\sum_{m\neq n}\frac{\langle \psi_{m}|ix^3|\psi_{n}\rangle}{E_n^{(0)}-
E_m^{(0)}}|\psi_{m}\rangle\nonumber  \\
&=&\frac{i}{2^{3/2}}\Big(\frac{1}{3}a^3 + 3\{a^2a^\dagger\} - 3\{a^{\dagger 2}a\}-\frac{1}{3}a^{\dagger 
3}\Big)|\psi_{n}\rangle\nonumber  \\
& = & \frac{i}{2^{3/2}} \Big(- \frac{2}{3} (a - a^{\dagger})^{3} +  \{(a+a^{\dagger})^{2} (a-a^{\dagger})\}
\Big)\nonumber\\
&=&-\left(\frac{2}{3}p^3+ x^{2}p -ix\right)|\psi_{n}^{(0)}\rangle\nonumber\\
&=&-\left(\frac{2}{3}p^3+ x^{2}p - ix\right)\sigma_{n}^{(0)}|\psi\rangle .
\end{eqnarray}
Thus, comparing with \eqref{R} we determine that to first order in perturbation
\begin{equation}
R = R^{(0)} + \epsilon R^{(1)} = \mathbbm{1} - \epsilon\left(\frac{2}{3}p^3+ x^{2}p - ix\right).
\end{equation}

Similarly, carrying out the perturbative calculation to order $\epsilon^3$ we determine (see 
\eqref{R}) 
\begin{equation}
R = \mathbbm{1} + \epsilon R^{(1)} + \epsilon^{2} R^{(2)} + \epsilon^{3} R^{(3)},
\end{equation}
where
\begin{eqnarray}
R^{(1) }& = & - \Big(\frac{2}{3}p^3+ x^{2}p - ix\Big),\nonumber\\
R^{(2)} & = & \Big(\frac{23}{288}p^6+ \frac{23}{96} x^{2}p^{4} - \frac{i}{48} xp^{3} + \frac{7}{96} 
x^{4}p^{2}\nonumber\\
& & \quad   - \frac{1}{16} p^{2} + \frac{13i}{48} x^{3}p + \frac{11}{8} x^{2} - \frac{41}{288} x^{6}\Big),
\nonumber  \\
R^{(3)} & = &\Big(\frac{59}{1296}\ p^9 +
  \frac{59}{288} \ x^2  p^7  - \frac{593i}{288}\ x p^6 -
\frac{163}{120}\ p^5\nonumber\\
& &\quad + \frac{109}{288}\ x^4  p^5 - \frac{1591i}{288}\  x^3 p^4 - \frac{769}{48}\ x^2  
p^3\nonumber\\
 & &\quad  +
  \frac{307}{864}\ x^6 p^3 +  \frac{649i}{48} \ x  p^2 - \frac{443i}{96}\ x^5
      p^2\nonumber\\
 & &\quad  - \frac{17}{12}p -13\ x^4p + \frac{41}{288}\ x^8p\nonumber\\
 & & \quad + \frac{685i}{72}x^3 - \frac{287i}{288}\ x^7\Big).
\end{eqnarray}
To this order, therefore, we can calculate
\begin{eqnarray}
R^{-1} & = & \mathbbm{1} - \epsilon R^{(1)} - \epsilon^{2} \left(R^{(2)} - (R^{(1)})^{2}\right)
\nonumber\\
&- & \!\!\!\! \epsilon^{3} \!\left(\!R^{(3)} + (R^{(1)})^{3} - R^{(1)}R^{(2)} - R^{(2)}R^{(1)}\!\right)\!.
\end{eqnarray} 
And from this we can calculate $q$ to order $\epsilon^{3}$ (see \eqref{qR}) to have the manifestly 
Hermitian form,
\begin{align}
q&= \mathbbm{1}-\epsilon\left(\frac{4}{3}p^3+2xpx\right)+\epsilon^2\left(\frac{169 }{144}p^6+
\frac{169 }{48}x p^4x\right.\nonumber\\
& \quad \left.+\frac{137 }{48}px^4 p+\frac{41 }{144}x^6-\frac{177 }{8}x^2\right)\nonumber\\
&\quad 
+\epsilon^3\left(  -\frac{181}{3} p-\frac{41}{36}   x^4 px^4+\frac{457 }{24}x^2 p x^2\right.\nonumber\\
&\quad -\frac{523 }
{108}x^3p^3 x^3+\frac{3463}{60} p^5 +\frac{5669}{72} xp^3x\nonumber\\
&\quad \left. -\frac{253}{36} x^2p^5x^2-\frac{155}{36}   x p^7x-\frac{155}{162}p^9
\right)+O(\epsilon^4).
\end{align}
This can be compared with \cite{bender-meisinger} to find linear order agreement. The difference in 
higher order terms is due to our choice of normalization $\langle \psi|\psi\rangle_{q} = \langle \psi|q|
\psi\rangle = 1$.

It is worth noting here that in this case the perturbing Hamiltonian is parity odd while the free 
harmonic oscillator Hamiltonian is invariant under parity (each term is individually ${\cal PT}$ 
invariant). As a result, the zeroth order energy eigenstates $|\psi_{n}^{(0)}\rangle$ are parity 
eigenstates and at every even order $|\psi_{n}^{(i)}\rangle$ would have the same parity as $|
\psi_{n}^{(0)}\rangle$ while at every odd order $|\psi_{n}^{(i)}\rangle$ would have the opposite 
parity. This leads to the fact that at odd order (where the correction to the energy can become 
imaginary)
\begin{equation}
E_{n}^{(i)} = 0,\quad i=2m+1.
\end{equation}
The even order corrections to energy, on the other hand, are all real. Therefore,
\begin{equation}
E_{n} = \sum_{i=0} \epsilon^{2i} E_{n}^{(2i)},
\end{equation}
is real at any order of perturbation. This is an alternative way of understanding the reality of energy 
in this model.

\subsection{Example: $H = \frac{1}{2} (p^{2} + x^{2} -1) + i\alpha x + i\epsilon x^{3}$}

From the previous results it is a simple matter to calculate $q$ in the case of the ${\cal PT}$ 
symmetric theory described by 
\begin{equation}
H=\frac{1}{2}(p^2+x^2-1) +i\alpha x+i\epsilon x^3,\label{alphaepsilon}
\end{equation}
where $\alpha,\epsilon$ are real constant parameters. First we note that with the canonical 
commutation relations we can write (see \eqref{harmonicoscillator})
\begin{eqnarray}
e^{-\alpha p/2}\,H_{\rm h.o.}\,e^{\alpha p/2}=H_{\rm h.o.} + i\alpha x-\alpha^2/4.
\end{eqnarray}
Therefore, in this case, we can write \eqref{alphaepsilon}
\begin{equation}
H = e^{-\alpha p/2}\, H_{\rm h.o.}\,e^{\alpha p/2} + \alpha^{2}/4 + i \epsilon x^{3} = H_{0} + i \epsilon 
x^{3},
\end{equation}
where we have identified
\begin{equation}
H_{0} = e^{-\alpha p/2}\,H_{\rm h.o.}\,e^{\alpha p/2} + \alpha^{2}/4.
\end{equation}
We can take this $H_{0}$ to be the free Hamiltonian in this case for which
\begin{equation}
E_{n}^{(0)} = E_{n}^{\rm (h.o.)} + \alpha^{2}/4,\quad |\psi_{n}^{(0)}\rangle = e^{-\alpha p/2} |
\psi_{n}^{\rm (h.o.)}\rangle.
\end{equation}
We can now carry over the perturbative analysis of the previous example on this state and determine $q$ (and, therefore, the inner product) to any order from 
\begin{eqnarray}
q=((Re^{-\alpha p/2})^{-1})^{\dagger}(Re^{-\alpha p/2})^{-1}=(R^{-1})^\dagger\;e^{\alpha p}\;R^{-1},
\end{eqnarray}
with $R$ calculated in the previous example. 

These two examples illustrate how the perturbative determination works when the exact energy 
eigenstates are difficult to obtain.

\section{Summary}

In this paper we have tried to build on our earlier proposal \cite{levi} for determining the positive inner product in the case of a pseudo-Hermitian Hamiltonian. The crucial concept in our method is the generator of energy eigenstates. We have clarified various aspects of our proposal which were not possible to explain in the earlier paper. We have shown through the example of the harmonic oscillator how our proposal reduces to the standard Dirac inner product when the Hamiltonian is Hermitian. We have given (additional) examples of finite dimensional pseudo-Hermitian matrix Hamiltonians to explain how our method works in practice. When the quantum mechanical Hamiltonian cannot be diagonalized exactly, the generator as well as the inner product need to be constructed perturbatively. We have developed the general formalism for this systematically for real energy eigenvalues. Furthermore, we have applied the formalism to two ${\cal PT}$ symmetric Hamiltonians to illustrate how it is carried out in practice.

\bigskip

\noindent{\bf Acknowledgments}
\medskip

This work was supported in part  by US DOE Grant number DE-FG 02-91ER40685,

\end{document}